\documentclass[10pt,twocolumn,conference]{IEEEtran}
\usepackage{amsfonts,amsmath,amsthm,latexsym,amssymb}
\usepackage{graphicx}
\usepackage{subfigure}
\usepackage{stfloats}
\usepackage{url}
\usepackage{multirow}
\usepackage[lined,boxed,commentsnumbered,ruled]{algorithm2e}
\begin{document}

\title{A Scalable Hybrid MAC Protocol for \\ Massive M2M Networks}
\renewcommand{\baselinestretch}{1}

\author{
\authorblockN{Yi~Liu$^\dag$, Chau~Yuen$^\dag$, Jiming Chen$^\ddag$,and Xianghui Cao$^\ddag$}
\authorblockA{ $\dag$Singapore University of Technology and Design,
Singapore\\
$\ddag$State Key Lab. of Industrial Control Technology, Dept. of
Control, Zhejiang University, Hangzhou, China
\\Email: \{yi\_liu, yuenchau\}@sutd.edu.sg, jmchen@ieee.org, xhcao@iipc.zju.edu.cn.}

}\maketitle

\begin{abstract}


In Machine to Machine (M2M) networks, a robust Medium Access Control
(MAC) protocol is crucial to enable numerous machine-type devices to
concurrently access the channel. Most literatures focus on
developing simplex (reservation or contention based) MAC protocols
which cannot provide a scalable solution for M2M networks with large
number of devices. In this paper, a frame-based Hybrid MAC scheme,
which consists of a contention period and a transmission period, is
proposed for M2M networks. In the proposed scheme, the devices
firstly contend the transmission opportunities during the contention
period, only the successful devices will be assigned a time slot for
transmission during the transmission period. To balance the tradeoff
between the contention and transmission period in each frame, an
optimization problem is formulated to maximize the system throughput
by finding the optimal contending probability during contention
period and optimal number of devices that can transmit during
transmission period. A practical hybrid MAC protocol is designed to
implement the proposed scheme. The analytical and simulation results
demonstrate the effectiveness of the proposed Hybrid MAC protocol.

\end{abstract}

\section{Introduction}
\IEEEPARstart{M}{a}{c}{h}{i}{n}{e}-to-Machine (M2M) communications
is defined as the information exchange between machines and machines
without any human interaction. With interconnection to the Internet
and deployed in different environments, a large number of devices
are autonomously organized to constitute an M2M network. M2M
networks are expected to be widely utilized in many fields of
pervasive applications \cite{CogM2M}, including industrial and
agricultural automation, health care, transport systems, electricity
grids, etc. There are two main characteristics of M2M networks: 1)
tremendous number of devices in service coverage and concurrent
network access attempt from these devices; 2) high level of system
automation in which the devices and systems can exchange and share
data. Therefore, the massive access management and medium access
protocol are the major issues in M2M communications to build up a
scalable, flexible, and automatic communication system
\cite{Forecast}.

Recently, the enormous economic benefits of the M2M communications
drive intensive discussion in international standardization
activities. In \cite{3GPP1}-\cite{802.16b}, 3GPP and IEEE studied
M2M requirements and possibilities. In order to handle the massive
access in M2M, 3GPP LTE has several work items defined on M2M
communications, primarily with respect to overload control
\cite{3GPP1} \cite{3GPP2}. IEEE 802.16p proposals \cite{802.16a}
\cite{802.16b} addressed enhancements for IEEE 802.16m standard to
support M2M applications. It is noted that the massive access
management of M2M communication over wireless channels generally
happen at the Medium Access Control (MAC) layer. Hence, the design
of a smart and efficient MAC protocol remains a key requirement for
successful deployment of any M2M networks.

As discussed by 3GPP and IEEE 802.16, the MAC protocol for M2M
communications focused on contention-based Random Access (RA)
schemes \cite{RA1} \cite{RA2} that allow all of the devices obtain
the transmission opportunities. The contention based RA is popular
due to its simplicity, flexibility and low overhead. Devices can
dynamically join or leave without extra operations. However, the
transmission collisions are eminent when huge number of M2M devices
trying to access the base station all at once. Reservation-based
schemes such as Time-Division Multiple Access (TDMA) \cite{TDMA}, is
well known as the collision-free access scheme. In this scheme, the
transmission time is divided into slots. Each devices transmits only
during its own time-slots. The main defects of TDMA is the low
transmission slot usage if only a small portion of devices have
information to transmit. Hybrid schemes attempt to combine the best
features of both of reservation-based and contention-based while
offsetting their weaknesses, e.g. \cite{hybrid1} \cite{hybrid2} try
to adapt to different bandwidth conditions depending on demand.

In this paper, we first propose a hybrid MAC protocol for M2M
networks, which will combine the benefit of both contention-based
and reservation-based protocols. In this scheme, each frame is
composed of two portions: Contention Only Period (COP) and
Transmission Only Period (TOP). The COP is based on CSMA/CA access
method, and is generally used for devices to contend for the
transmission slots in TOP. Only successful contending devices are
allowed to transmit data during the TOP that provides TDMA type of
data communication. Given the frame duration, it is expected that
the number of the successful devices increases when the COP duration
is prolonged. However, the COP duration increases at the expense of
shortening the TOP which results in the decrease of transmission
slots. To achieve the optimal tradeoff between the contention and
transmission period in each frame, an optimization problem is
formulated to maximize the throughput by deciding the optimal
contending probability during COP, and the optimal number of devices
to transmit during TOP (which is related to the duration of COP).
Then, we design a hybrid MAC protocol to implement the proposed
scheme in a practical environment. The analytical and simulation
results demonstrate the effectiveness of the proposed hybrid MAC
protocol.
%
%
%
%
%
%
%
%
%

The remainder of this paper is organized as follows. In Section II,
we describe the system model of a M2M network. Then, we propose a
hybrid access control scheme by optimizing the duration of COP and
TOP in Section III. In Section IV, we design a hybrid MAC protocol
to implement the proposed scheme. Performance study and evaluation
are given in Section V. Section VI concludes the paper.

\section{System Model}
We consider a M2M networks which consists of one base station (BS)
and $K$ number of devices. BS dominates medium access control for
all the devices and there are $L$ number of devices that have data
to transmit during one frame (the values of $K$ and $L$ may vary
from frame to frame). Hence in every time frame, there are $K-L$
silent devices that can put into sleep mode as they have no data to
transmit. In our system model, we assume that a homogeneous scenario
where all of the devices have same amount of data with same
priority. Hence, each device has to contend the transmission
opportunities when it has data to transmit. We assume there $M$
devices succeed in COP, and secure a transmission time slot in TOP.
The fairness for different types of M2M devices with different QoS
requirement will be our future work.

\begin{figure}[htb]
\centerline{\includegraphics[width=9cm]{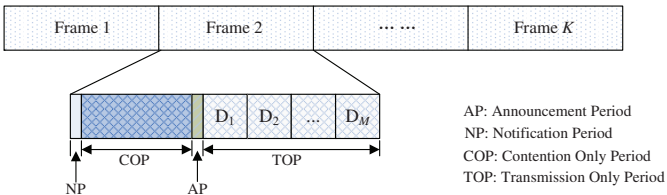}}
\caption{The frame structure.} \label{frame}
\end{figure}

To show the universality of our model, the basic timing unit of the
access operation is the frame which is composed of four portions as
depicted in Fig. \ref{frame}: Notification Period (NP), Contention
Only Period (COP), Announcement Period (AP) and Transmission Only
Period (TOP). The BS broadcasts notification messages at NP to all
devices to notify the beginning of the contention. The $L$ devices
has data to transmit will contend the channel during COP. The COP is
based on $p$-persistent CSMA access method \cite{p-persistent}, and
is generally used for devices to randomly send the transmission
requests to BS. The remaining time of a frame is specified as the
TOP, which provides TDMA type of communication for the $M$ devices
which successfully contend the transmission slots. Without loss of
generality, we assume that each assigned transmission slot has the
same length.

We note that the BS will assign the transmission slots to the
devices which successfully contend the transmission slots in COP. It
is expected that higher number of successful devices ($M$) can be
obtained if $T_{COP}$ get longer. However, given the duration of a
frame $T_{frame}$, the $T_{TOP}$ will be decreased, which may reduce
the transmission time for the successful devices. Hence, there is a
tradeoff between the duration of COP and TOP. To balance this
tradeoff, we attend to propose a hybrid access control scheme which
focus on maximizing the aggregate throughput by finding the optimal
duration of COP and TOP in the next section.

\section{Hybrid Access Control Scheme}
In this section, for a given the number of the contending devices
$L$ and the duration of a frame $T_{frame}$, we derive the optimal
contending probability for all the contending $L$ devices during COP
(denoted as $p_{opt}$), and the optimal number of devices that can
transmit during TOP (denoted as $M_{opt}$).

Based on $p$-persistent CSMA in COP period, when a contention
attempt is completed (successfully or with a collision), each
contending device (i.e. the device with packets ready for
transmission) will start a contention attempt with probability $p$.
Here, we define the successful contention as the event that the
transmission request from a device is successfully received by BS.
Let $t_i$ denote the time between the ($i-1$)th and the $i$th
successful contention. Considering the behavior of the
$p$-persistent CSMA, we let $N_i^c$ denote the number of collisions
that occur during $t_i$, then
\begin{equation}
t_i=\sum_{j=1}^{N_i^c}[\mathrm{Idle}_{i,j}+\mathrm{Coll}_{i,j}]+\mathrm{Idle}_{N_i^c+1}+S_i
\end{equation}
where $\mathrm{Idle}_{i,j}$ is the duration of the $j$th idle time
that precedes the channel busy period (either collision or success)
in each $t_i$ duration. $\mathrm{Coll}_{i,j}$ is the duration of the
$j$th collision given that a collision occurs, and $S_i$ is the
length of the request message. Let $T_{COP}$ denote the duration of
the COP in each frame. Then, we have
\begin{eqnarray}
T_{COP}\!\!\!\!\!\!\!\!\! && =\sum_{i=1}^M t_i\\
&&
=\sum_{i=1}^M\!\!\left\{\sum_{j=1}^{N_i^c}[\mathrm{Idle}_{i,j}+\mathrm{Coll}_{i,j}]+\mathrm{Idle}_{N_i^c+1}+S_i\!\!\right\}.
\end{eqnarray}
Since $T_{COP}$ is the sum of random variable $t_i, (i=1,\cdots,
M)$, the $T_{COP}$ is also a random variable with $E[T_{COP}]$ the
average time for $M$ number of successful contentions. To obtain the
close-form expression of the $T_{COP}$, we then focus on deriving
the $E[T_{COP}]$ in order to determine $M_{opt}$ and $p_{opt}$. Due
to the independently distributed $t_i$, we have
\begin{equation}\label{TCOP}
\begin{split} \quad & E[T_{COP}] =  \sum_{i=1}^M E[t_i]\\
&
=\sum_{i=1}^M\left\{(E[{N_i^c}]+1)E[\mathrm{Idle}_{i}]+E[{N_i^c}]E[\mathrm{Coll}_{i}]+E[S_i]\right\}
\end{split}
\end{equation}
where $E[{N_i^c}]$, $E[\mathrm{Idle}_{i}]$, $E[\mathrm{Coll}_{i}]$
and $E[S_i]$ are the average number of collisions, the average
duration of a idle time, a collision and an request message during
$t_i$, respectively. According to \cite{p-persistent}, we have the
following expressions:
\begin{equation}
\begin{split} & E[{N_i^c}] =
\frac{1-(1-p)^{L-i}}{(L-i)p(1-p)^{L-i-1}}-1\\ \nonumber &
E[\mathrm{Idle}_{i}]=\frac{(1-p)^{L-i}}{1-(1-p)^{L-i}}\cdot
\delta_{idle}
\end{split}
\end{equation}
where $\delta_{idle}$, $\delta_{coll}=E[\mathrm{Coll}_{i}]$ and $
\delta_{succ} = E[S_i]$ are constant \cite{p-persistent}. Then,
$E[T_{COP}]$ is the function of $M$ and $p$. Let
$\mathcal{T}_{COP}(M,p)=E[T_{COP}]$, after some algebraic
manipulations:
\begin{equation}\label{TCOPf}
\begin{split} \mathcal{T}_{COP}(M,p) & =
\sum_{i=1}^M\left\{\frac{(1-p)^{L-i}}{(L-i)p(1-p)^{L-i-1}} \cdot
\delta_{idle} \right. \\
&  \left.+ \left( \frac{1-(1-p)^{L-i}}{(L-i)p(1-p)^{L-i-1}}-1
\right) \cdot \delta_{coll} + \delta_{succ} \right\}.
\end{split}
\end{equation}

Given $T_{frame}$, longer $\mathcal{T}_{COP}(M,p)$ allow more
devices succeed in contention. However, the incremental
$\mathcal{T}_{COP}(M,p)$ will reduce the duration of TOP subjecting
to the constraint as $\mathcal{T}_{COP}(M,p)+T_{TOP}\leq T_{frame}$.
To balance this tradeoff, we formulate an optimization problem to
maximize the aggregate throughput in each frame. Here, the aggregate
throughput is defined as the sum of the throughput obtained by all
the devices which are allocated the transmission slots during each
frame. Let $T_{tran}$ and $R$ denote the the transmission time slot
and data rate of each device which are constant. Then, we can
maximize the aggregate throughput, denoted by $C_{total}$, for each
frame as
\begin{eqnarray}\label{Optimal}
\{M_{opt}, p_{opt}\} = && \max_{M,p}  \quad C_{total} = \max_{M,p}
\quad  M R T_{tran}
\\ \quad
s.t.   \quad && \mathcal{T}_{COP}(M,p)  +   M T_{tran} \leq T_{frame} \\
       \quad \quad \quad && 0  \leq   p  \leq 1
\end{eqnarray}
Then, we try to prove the convexity of above optimization problem.
Since the objective function in (\ref{Optimal}) is a convex function
of $M$ and constraint (8) is linear. Lemma \ref{Lemma} below shows
that, asymptotically, for M2M networks with tremendous number of
devices, i.e. $L$ is large, the constraint (7) is also a convex
function.

\newtheorem{theorem}{Lemma}
\begin{theorem}\label{Lemma}
For $L\rightarrow\infty$, $\mathcal{T}_{COP}(M,p)$ can be obtained
as a convex function of $M$ and $p$.
\end{theorem}

\begin{IEEEproof}
Since the duration of $T_{frame}$ has a finite value, as
$L\rightarrow\infty$, it is easy to obtain $L \gg M$, then we have
\begin{equation}
\begin{split}  \mathcal{T}_{COP}(M,p) & =
M \cdot \left\{\frac{(1-p)^{L}}{Lp(1-p)^{L-1}} \cdot
\delta_{idle} \right. \\
&  \left.+ \left( \frac{1-(1-p)^{L}}{Lp(1-p)^{L-1}}-1 \right) \cdot
\delta_{coll} + \delta_{succ} \right\}
\end{split}
\end{equation}
Moreover, $(1-p)^{L-1}$ tends to $(1-p)^{L}$ if $L$ sufficiently
large. Hence, we can obtain the approximated transformation of the
above equation as
\begin{equation}\label{optimization}
\begin{split}  \mathcal{T}_{COP}(M,p) & =
M  \left\{\frac{1}{Lp}  \delta_{idle}  + \delta_{succ} \right. \\
& \left.  + \left( \frac{1}{Lp(1-p)^{L-1}} - \frac{1}{Lp} - 1\right)
\delta_{coll} \right\}
\end{split}
\end{equation}

Taking the second derivative of $\mathcal{T}_{COP}(M,p)$ with
respect to $M$ and $p$, respectively, gives
\begin{equation}
\begin{split}\notag
\frac{\partial^2\mathcal{T}(M,p)}{\partial M^2} & =0\\
\frac{\partial^2\mathcal{T}(M,p)}{\partial p^2} & =\frac{2}{Lp^3}
\delta_{idle}  + \left( \frac{1+(1-p)^{L+1}+Lp}{p^2(1-p)^{L+2}}
 \right) \delta_{coll} > 0
\end{split}
\end{equation}
Consequently, we conclude that $\mathcal{T}_{COP}(M,p)$ is a convex
function of $M$ and $p$ \cite{Optimal}.
\end{IEEEproof}

Therefore, the optimization problem is a convex programming problem
and can be solved easily with off-the-shelf toolbox. And the optimal
period of COP, $T_{COP,opt} = T_{COP}(M_{opt},p_{opt})$.

\section{A Practical Hybrid MAC Protocol Design}

In this section, we design a practical hybrid MAC protocol for the
M2M networks. The operations of the proposed hybrid MAC are
separated into frames for contention and data transmission as shown
in Fig. \ref{frame}. As mentioned early, each frame is divided into
four periods: NP, COP, TOP, and AP. The specific description of each
period is given as follows:

\subsection{Notification Period (NP)}
At the start of every time frame, the BS broadcasts an advertisement
message (ADV) to all $K$ number of devices. BS then estimate the
number of devices that have information to transmit (i.e. value of
$L$), one way is by using the estimation method proposed in
\cite{hybrid1}. Then, the BS broadcast the duration of contention
period $T_{COP,opt}$ and the contending probability $p_{opt}$ based
on optimization solution in Section III. And the networks enters
COP.

\subsection{Contention Only Period (COP)}

In this period, $L$ number of the devices contend the transmission
opportunities based on $p$-persistent CSMA method, with contending
probability of $p_{opt}$. The contending devices randomly send the
transmission request (Tran-REQ) message to the BS. The contention is
declared as success only when one device send the Tran-REQ message.
When more than one devices are sending Tran-REQ, the collision
occurs. The idle period is a time interval in which the contention
is not happening. Under $p$-persistent CSMA, the success period and
collision period can be given as
$\delta_{coll}=E[\mathrm{Coll}_{i}]=T_{req} + BIFS$ and $
\delta_{succ} = E[S_i] = T_{req} + SIFS + T_{ACK} + BIFS$, where
$T_{req}$ is the length of Tran-REQ message, $T_{ACK}$ is the
duration of ACK, and $BIFS$ and $SIFS$ are the backoff inter frame
space and short inter frame space respectively.

\begin{figure}[htb]
\centerline{\includegraphics[width=7cm]{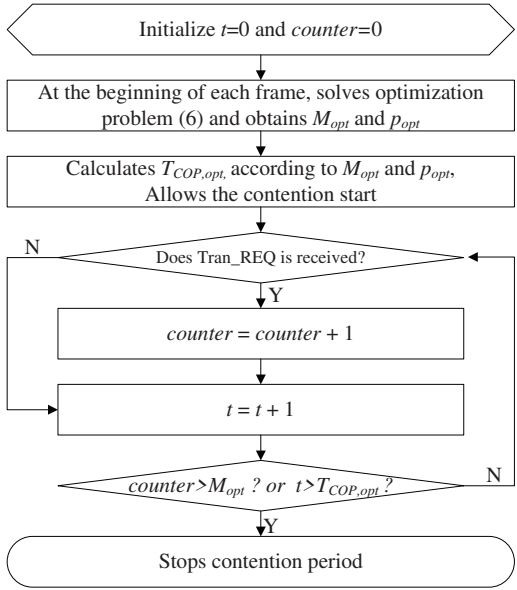}}
\caption{A flowchart of the two thresholds scheme.} \label{workflow}
\end{figure}

If a Tran-REQ message successfully received, the BS increments the
counter by one. To control the number of successful devices, the
optimal $M_{opt}$ which obtained by (\ref{optimization}) is used as
the threshold. Recall that the calculation of $M_{opt}$  in
(\ref{optimization}) is based on the expected value calculation
which is not able to manifest the variability of the devices' random
contention. To avoid performance degradation caused by the
difference between the analytical and practical results, we propose
a two thresholds scheme to control the duration of COP as shown in
Fig. \ref{workflow}. In this scheme, the BS stop the COP period not
only depending on the number of successful devices $M$, but also the
contention period $\mathcal{T}_{COP}$. Either the counter value
exceed $M_{opt}$  or the real contention time longer than
$\mathcal{T}_{COP,opt}$, the BS will stop the COP and declare the
next period, i.e., announcement period. Hence there will be at most
$M_{opt}$ devices has the right to transmit during TOP.

\subsection{Announcement period (AP)}

After the contention finished, BS initiates and broadcasts the
announcement message to all of the contending devices. The
announcement message consists of two parts: (i) successful devices'
ID and (ii) the transmission schedule. If the device verify its' ID
in the message, the device should send data at the assigned
transmission slots. If the device do not verify its' ID, the device
should go into the sleep mode and wait for the next frame. Such
arrangement keep the wake-up time of a device at minimal, and we
will further investigate the energy consumption of the proposed
schemes in the future work.

\subsection{Transmission Only Period (TOP)}

In the data transmission period, the successful device turns on its
radio module and sends its data to the BS over its allocated time
slots, and turns its radio module off at all other times. Though
only uplink is mentioned, some modification to the protocol can be
applied to downlink where the devices would like to receive
information from the BS. While we only focus on homogenous scenario,
we will extend the work further by considering each device has a
different priority, and different data size.

\section{Performance Study and Evaluation}
In the following sections, we will compare the proposed hybrid MAC
procotol with contention-based protocol - slotted Aloha \cite{RA2}
and reservation-based protocol - TDMA \cite{TDMA} in terms of
throughput, utility and average transmission delay. The simulation
parameters are shown in Table \ref{table1}.

In Table \ref{table2}, we present $M_{opt}$ and $p_{opt}$ in terms
of the duration of the frame $T_{frame}$, when the number of the
contending devices $L$ equals to 100, 200 and 300, respectively. It
is observed that $p$ decreases as $L$ increases which shows the
overload control ability of our scheme. Moreover, the number of the
successful devices $M_{opt}$ is double if $T_{frame}$ is doubled.
This indicates that our scheme is very efficient, as we can set
$T_{frame}$ to be small, and it works well for any values of $L$
without any lost in efficiency.

\begin{table}[!t]
\renewcommand{\arraystretch}{1.2}
\caption{The simulation parameters} \label{table1} \centering
\begin{tabular}{|c||c||c|}
\hline
$T_{NP}$ & 10.2 $\mu$s & The duration of NP\\
\hline
$T_{AP}$ & 10.2 $\mu$s & The duration of AP\\
\hline
$T_{tran}$ & 1 ms & The transmission time of each device\\
\hline
$R$ & 1.728 Gbps &  The data rate\\
\hline
$T_{req}$ & 22.2 $\mu$s & The length of Tran-REQ message\\
\hline
$T_{ACK}$ & 7.5 $\mu$s & The duration of ACK frame\\
\hline
$SIFS$ & 2.5 $\mu$s & The duration of short interframe spacing\\
\hline
$BIFS$ & 7.5 $\mu$s & The duration of backoff interframe spacing\\
\hline
\end{tabular}
\end{table}

\begin{table}[!t]
\renewcommand{\arraystretch}{1.2}
\caption{The optimization results} \label{table2} \centering
\begin{tabular}{|c||c||c||c||c||c|}
\hline
\multicolumn{3}{|c||}{$T_{frame}=50 ms$} &  \multicolumn{3}{|c|}{$T_{frame}=100 ms$}\\
\hline
$L$ & $M_{opt}$ & $p_{opt}$ & $L$ & $M_{opt}$ & $p_{opt}$\\
\hline
$100$ & 46  & 0.06 & $100$ & 92 &  0.06\\
\hline
$200$ & 46  & 0.03 & $200$ & 92  & 0.03\\
\hline
$300$ & 46  & 0.02 & $300$ & 92  & 0.02\\
\hline
\end{tabular}
\end{table}

\subsection{Throughput}
\begin{figure}[htb]
\centerline{\includegraphics[width=8cm]{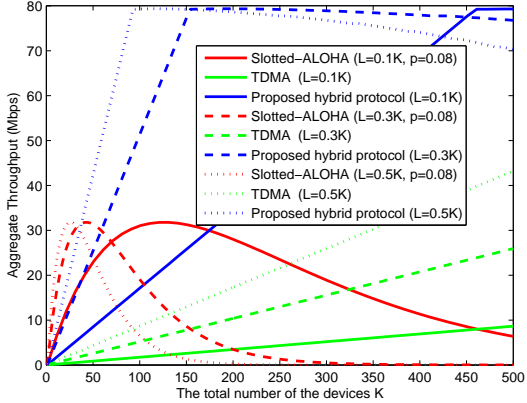}}
\caption{The aggregate throughput comparison in terms of the total
number of the devices when $T_{frame}=50ms$.} \label{TH_Com}
\end{figure}

Then, we compare the aggregate throughput in terms of the total
number of the devices ($K$). Here, the transmission probability in
slotted-ALOHA is set as 0.08. Under such setting-up, the
relationship between $K$ and $L$ is $L=10\% K$, $L=30\% K$ and
$L=50\% K$. As shown in Fig. \ref{TH_Com}, the aggregate throughput
in the proposed hybrid protocol is always higher than TDMA, and will
higher than slotted-ALOHA as $K$ and $L$ increase. That is, the
proposed hybrid protocol can optimally control the contention
probability $p$ and the number of successful devices to maximize the
aggregate throughout. While slotted-ALOHA performs well only at
low-load condition, and TDMA performs well only at heavy-load
condition. While our hybrid scheme may not perform the best under
low-load condition (e.g. $L=10\%K$ and $K$ is small), our hybrid
scheme outperform the other two when $L$ is increased.

\begin{figure}[htb]
\subfigure[]{\begin{minipage}{0.5\linewidth}
{\includegraphics[width=1.9in]{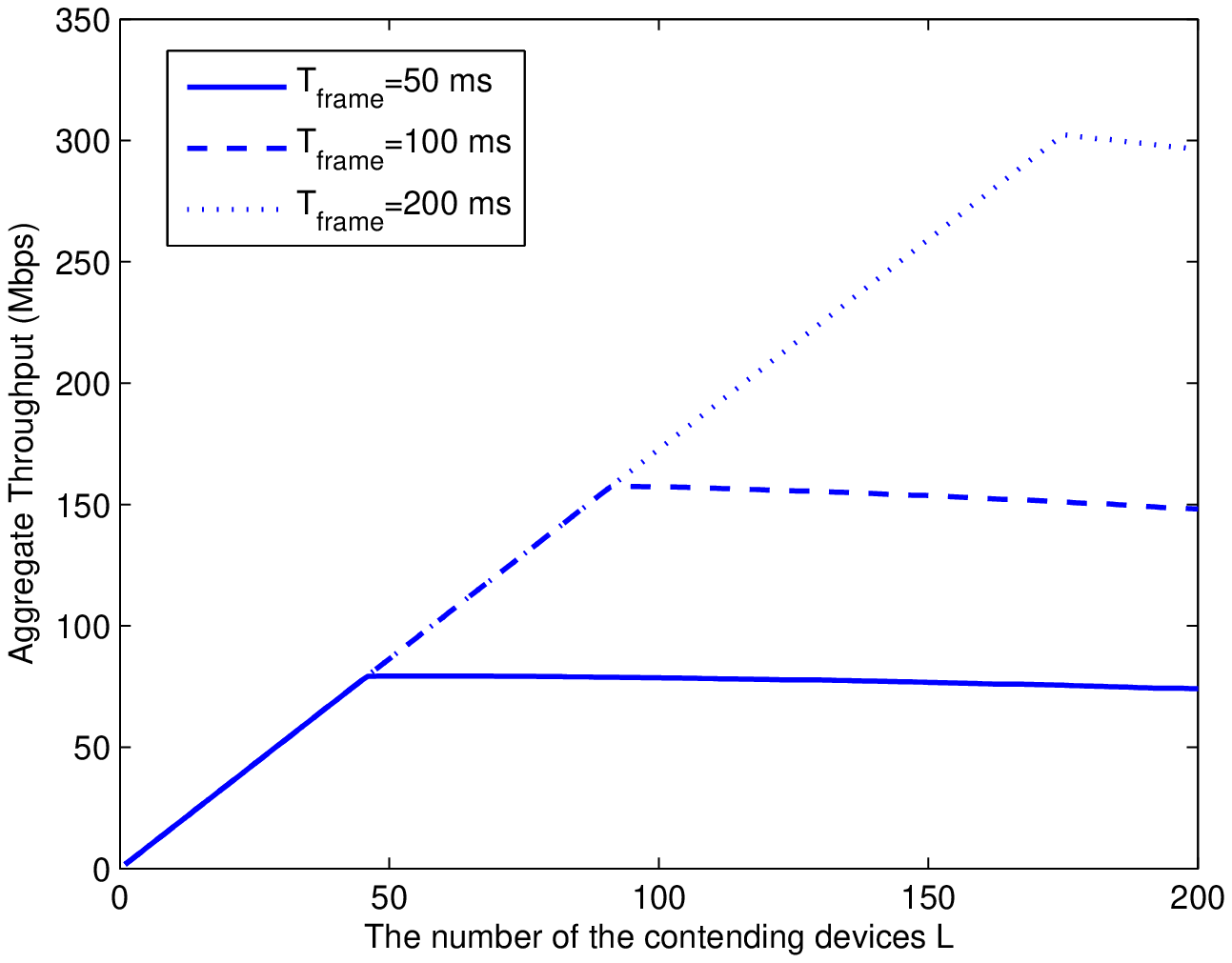}}
\end{minipage}}\subfigure[]{\begin{minipage}{0.5\linewidth}
{\includegraphics[width=1.9in]{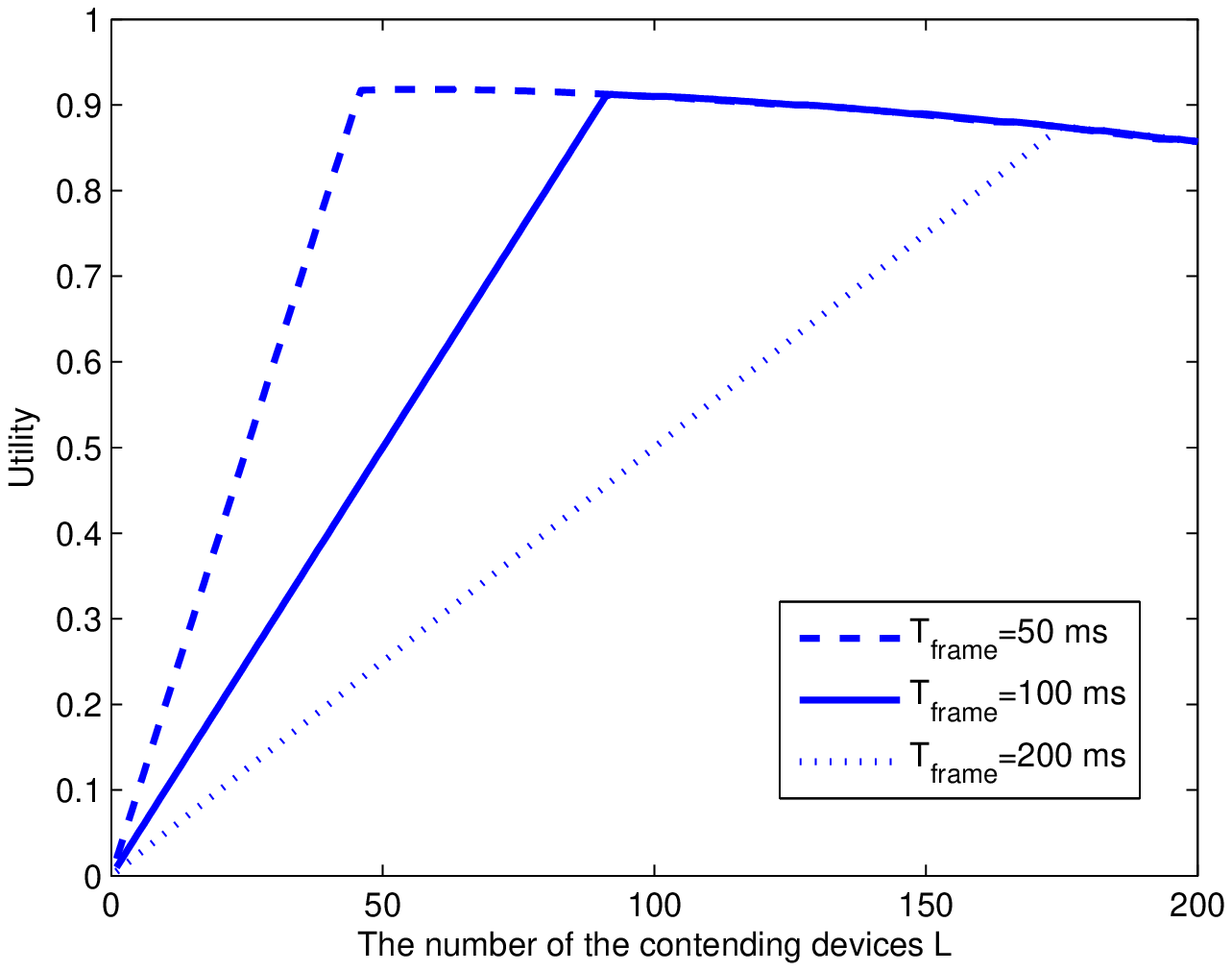}}
\end{minipage}}
\caption{The aggregate throughput and utility in terms of the number
of the contending devices.}  \label{THM_L}
\end{figure}

The aggregate throughput in terms of the number of the contending
devices $L$ for different value of the $T_{frame}$ is shown in Fig.
\ref{THM_L}(a). For a given $T_{frame}$, it is observed that the
throughput linearly increase at first as the number of the
contending devices increases until the maximal throughput is
obtained. Then, the aggregate throughput has a slight drop after the
number of contending device $L$ exceed $M_{opt}$.


\subsection{Utility Comparison}

In this paper, we define the utility as the ratio of transmission
period ($T_{TOP}$) to the period of each frame ($T_{frame}$). Let
$U$ denote the utility, we have
\begin{equation}
U=\frac{T_{TOP}}{T_{frame}}
\end{equation}
To illustrate the performance of utility, we also compare our
proposed protocol with the slotted-ALOHA and TDMA. Similarly, the
transmission probability in slotted-ALOHA is set as 0.08.

\begin{figure}[htb]
\centerline{\includegraphics[width=8cm]{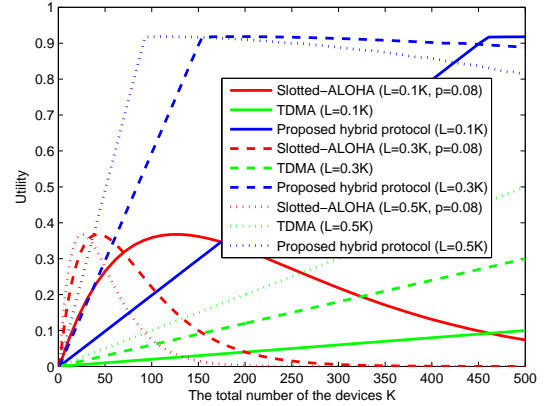}} \caption{The
utility in terms of the number of the total number of the devices
when $T_{frame}=50ms$.} \label{U_compare}
\end{figure}

Fig. \ref{U_compare} shows the utility in terms of the total number
of devices when $T_{frame}=50ms$. Under such setting-up, the
relationship between $K$ and $L$ is $L=10\% K$, $L=30\% K$ and
$L=50\% K$. We observe that the utility of the proposed hybrid
protocol is lower than that in slotted-ALOHA when the number of the
contending devices is low. However, as the number of the devices
increases, the collision caused by slotted-ALOHA will increase which
can drastically reduce the utility. Since the hybrid protocol use
the TDMA mechanism for transmission, the successful devices can
transmit data without collision. In addition, the utility in
proposed hybrid protocol is especially higher than that in TDMA.
This is because the proposed hybrid protocol only allow the devices
with data to transmit to participate in the contention. Hence, the
transmission slots assigned to the successful devices can be fully
utilized. Comparatively, the slots assignment in TDMA is fixed and
static for each device without considering the thorough channel
utilization.

Fig. \ref{THM_L}(b) shows the utility in terms of the number of
contending devices. For a given $T_{frame}$, it is observed that the
utility linearly increase at first and then decrease as the number
of the contending devices increases. Together with the results in
Table \ref{table2} and Fig. \ref{THM_L}, we can set the $T_{frame}$
to be some small value, and yet it can achieve the same normalized
throughput / utility as high value of $T_{frame}$. This eliminate
the worry about the value of contending devices $L$ when setting
$T_{frame}$, as in practical network $L$ may not known before hand.

\subsection{Average Transmission Delay}
In this subsection, we aim to compare the average transmission delay
among the proposed hybrid protocol, slotted-ALOHA and TDMA. The
transmission delay is defined as the time elapsed between the start
of a frame and the end of its transmission to the BS during a frame.
Without loss of generality, the transmission probability in
slotted-ALOHA is set as 0.08, and $T_{frame}=200ms$. Meantime, the
relationship between $K$ and $L$ is $L=10\% K$, $L=30\% K$ and
$L=50\% K$. From \cite{p-persistent} and \cite{TDMA}, it is easy to
obtain the average transmission delay in slotted-ALOHA and TDMA.
Then, we focus on evaluating the average transmission delay for the
proposed hybrid protocol. Let $T_{delay}$ denote the average
transmission delay, we have
\begin{equation}
T_{delay}=T_{NP}+\overline{T}_{COP}+T_{AP}+\overline{T}_{TOP}
\nonumber
\end{equation}
where $\overline{T}_{COP}=\mathcal{T}_{COP}$. After contention, the
BS initiates and broadcasts a transmission schedule for the
successful devices. After receiving the schedule, each successful
device sends its data packet to the BS at its scheduled time slots
$T_{tran}$ following TDMA mechanism. Hence, the average delay for a
device during a single frame is $\overline{T}_{TOP} =
\frac{T_{frame}-\mathcal{T}_{COP}-T_{NP}-T_{AP}}{2}\left(1-\frac{1}{M_{opt}}\right)+T_{tran}$.

\begin{figure}[htb]
\centerline{\includegraphics[width=8cm]{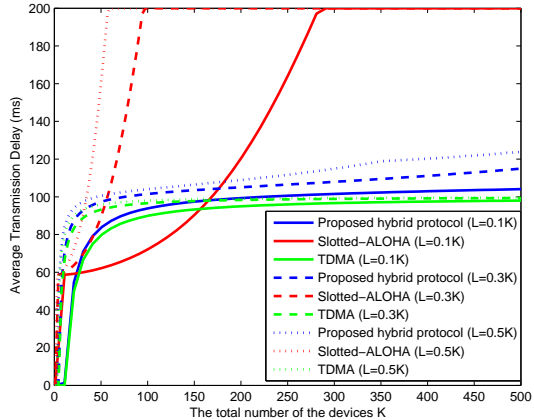}}
\caption{The comparison of average transmission delay in terms of
the total number of devices when $T_{frame}=200ms$.} \label{delay}
\end{figure}

Fig.\ref{delay} shows the average transmission delay comparison
among Slotted-ALOHA, TDMA and proposed hybrid protocol. The
comparison indicates that hybrid protocol is able to achieve lower
delay than Slotted-ALOHA. This is because hybrid protocol only allow
the device transmit a small command message during contention
period. The waiting time during collision can be greatly reduced.
Moreover, when the number of the devices becomes large, our hybrid
scheme not only control the number of the served devices but also
control the transmission probability to mitigate the congestion of
the devices. In addition, proposed scheme has litter higher but
closed results compared to the TDMA scheme. That is, in the proposed
scheme, devices have to spend more waiting time during the
contention period, however, this is the trade-off to achieve a
higher utility of channel.

\section{Conclusion}
In this paper, we focused on studying the massive access control
scheme for M2M networks. In our scheme, the operation of each frame
is mainly divided into two parts: contention only period (COP) and
transmission only period (TOP). The devices only send contending
commands during COP and transmit data during TOP. Under such
mechanism, the BS can easily maximize the aggregate throughput by
controlling the duration of COP and TOP which are decided by the
contending probability of the devices and the number of the served
devices. An optimization is formulated to solve the problem, and we
show analytically the problem is convex. To implement the scheme, we
then presented a hybrid MAC protocol for the M2M networks. We
analyzed the aggregate throughput, utility and the average
transmission delay to show the effectiveness of the proposed hybrid
MAC protocol.

In the future, we will consider the fairness for our proposed scheme
and in a more practical environment: heterogeneous M2M network where
the devices may have different service requirements. In order to
fairly assign the resources to these devices, a QoS provisioning
access control scheme should be considered. In that case, new
constraints should be added in the optimization problem to cover the
heterogenous among all type of devices.


\section*{Acknowledgment}
\addcontentsline{toc}{section}{Acknowledgment}

This research is partly supported by the Singapore University
Technology and Design (grant no. SUTD-ZJU/RES/02/2011), NRF EIRP,
the Natural Science Foundation of China (grant no. 61203036) and
China postdoctoral special funding (grant no. 2012T50516).

\appendices

%




\end{document}